\documentclass{aa}  
\usepackage{graphicx}
\usepackage{txfonts}
   \newcommand{\unit}[1]{\ensuremath{\, \mathrm{#1}}}
   \newcommand{\ramses}{{\sc ramses}}          
   \newcommand{\Fig}[1]{Fig.~\ref{fig:#1}}    
   \newcommand{\Figure}[1]{Figure~\ref{fig:#1}}    

\begin{document}

   \title{Constraints on the (re-)orientation of star-disk systems through infall}

   \subtitle{}

   \author{M. Kuffmeier
          \inst{1}
          \and
          J. E. Pineda\inst{2}
          \and
          D. Segura-Cox \inst{3}
          \and
          T. Haugb{\o}lle \inst{1}
          }
   \institute{Niels Bohr Institute, University of Copenhagen, {\O}ster Voldgade 5, DK-1350 Copenhagen, Denmark
               \and
              Max-Planck Institute for Extraterrestrial Physics, Gie{\ss}enbachstra{\ss}e 1, 85748 Garching               \email{kueffmeier@nbi.ku.dk}          
               \and
   Department of Astronomy, The University of Texas at Austin, 2515 Speedway, Stop C1400 Austin, Texas 78712-1205, USA
}

   \date{Received April 16, 2024}

 
  \abstract
    {
    It has been consensus for decades that star-disk systems accrete most of their mass and thereby their angular momentum during the collapse of a prestellar core, such that the rotational direction of an individual system is equivalent to the net rotation of the core.
    Recent results, however, indicate that stars experience events of post-collapse infall (often referred to as late infall), during which the star and its disk is refreshed with material from the protostellar environment through accretion streamers. 
    Apart from adding mass to the star-disk system, infalling material potentially supplies a substantial amount of angular momentum to the system as the infalling material is initially not bound to the collapsing prestellar core.
    In this paper, we investigate the orientation of infall on star-disk systems by analyzing the properties of accreting tracer particles in 3D magnetohydrodynamical simulations of a molecular cloud that is (4 pc)$^3$ in volume. The simulations were carried out with the adaptive mesh refinement code \ramses. 
    In contrast to the traditional picture, where a star-disk system inherits its rotational axis from the collapse of a coherent pre-stellar core, the orientation of star-disk systems can change substantially during the accretion process. 
    In agreement with previous results that show larger contributions of late infall for increasing stellar masses, infall is more likely to lead to a prolonged change in orientation for stars of higher final mass.
    While the average change in orientation varies significantly between stars with different final masses, the change in orientation over the relative mass accrual per star is similar, independent of the final mass.    
    The probability for stars to accrete their mass with another star in their $\approx 10000$ au surrounding is larger than accreting as a single star.
    On average, brown dwarfs and very low mass stars are more likely to form and accrete all of their mass as part of a multiple system, while stars with final masses above a few 0.1 M$_{\odot}$ are more likely to accrete part of their mass as single stars. 
    Finally, we find an overall trend: the post-collapse accretion phase is more anisotropic than the early collapse phase. 
    This result is consistent with a scenario, where mass accretion from infall occurs via infalling streamers along a preferred direction, while the initial collapse is less anisotropic albeit the material is funneled through accretion channels.  
    }


   \maketitle
%

\section{Introduction}
Stars form from the gravitational collapse of dense gas in Giant Molecular Clouds. 
Traditionally, the key properties of a star (such as its mass and angular momentum) are understood as the outcome of the spherical collapse of an accumulation of gravitationally bound gas (`prestellar core') of the order of mass-dependent free-fall times \citep{Larson1969,Shu1977,Terebey+1984}
\begin{equation}
    t_{\rm ff} = \sqrt{\frac{\pi^2 R^3}{8 G M}},
\end{equation}
where $R$ is the radius of the collapsing core, $G$ is the gravitational constant and $M$ is the mass of the core. 
The protostar forms together with a disk as a result of angular momentum conservation and gains most of its mass at the beginning of the collapse sequence, while the main accretion phase only lasts for, at most, up to a few 100 kyr \citep{Zhao+2020}. 
Initially, the star-disk system is deeply embedded in its envelope; the remnant material of the collapsing prestellar core. 
Over time, the remaining material mildly rains from the envelope onto the star-disk system and the star-disk system becomes less embedded. 
The spectral energy distributions of a protostar can be used to determine how embedded the protostar is, and it is hence used as a tracer of the evolutionary stage of protostars \citep{LadaWilking1984,Lada1987,Andre+1993}.

The understanding of star formation is currently undergoing a paradigm change though. 
It changes from a picture where the properties of a star and its disk are solely inherited from a collapsing prestellar core towards a picture where the properties of star-disk systems can be significantly altered by external effects after the initial collapse phase.
In clustered regions, massive stars can modify the properties of disks around low-mass stars via external photoevaporation leading to shrinking of the disk and disk dispersal \citep[see][and references therein]{WinterHaworth2022}. 
Apart from that, observations of filamentary structures associated with protostars, so-called streamers, strongly suggest interactions with the protostellar environment beyond the initial collapse phase \citep{Pineda+2019,Pineda+2023,Yen+2019,Alves+2020,Ginski+2021,Huang+2021,Garufi+2021,Valdivia-Mena+2022,Valdivia-Mena+2023,Valdivia-Mena+2024,Gupta+2023,Cacciapuoti+2024,Zurlo+2024}.
One possible explanation for some of these structures are interactions with other stars, so called stellar fly-bys that affect the disk and its properties \citep[see][and references therein]{Cuello+2023}.
Another explanation is the possibility of encountering material from the Giant Molecular Cloud environment that was gravitationally not bound to the prestellar core \citep{Pelkonen+2021}. In this way, the star-disk system can be fed with substantial amounts of fresh material after the initial collapse phase. 
This latter scenario is referred to as `post-collapse infall' or  `late infall' \citep{Kuffmeier+2023}. 
Such events can lead to a significant increase in the protostellar accretion rate, possibly explaining luminosity bursts \citep{Padoan+2014,JensenHaugbolle2018,Kuffmeier+2018}, apparent rejuvenation of the protostar when it becomes more embedded again during an infall event \citep{Kuffmeier+2023}, and even to the formation of a second-generation disk \citep{Thies+2011,Kuffmeier+2020} that is likely misaligned with respect to the primordial disk \citep{Bate2018,Kuffmeier+2021}. 
Intriguingly, misalignment between inner and outer disk is also the best explanation for the observation of shadows in protoplanetary disks \citep{Marino+2015}.

Interestingly, several works showed that the relative amount and frequency of post-collapse infall increases with increasing final mass of the star \citep{Smith+2011,Padoan+2020,Kuffmeier+2023}.
This implies that stars can accrete a substantial amount of mass after their initial collapse phase, which can explain the absence of massive cores in recent observational surveys \citep{Sanhueza+2019,Morii+2023,Cheng+2024}.
As first shown by \cite{Pelkonen+2021}, even solar-mass stars accrete \textit{on average} almost $50 \%$ of their mass during this second phase.
Some stars accrete almost all of their mass during the initial collapse phase and experience only negligible post-collapse infall, but others gain most of their mass during the post-collapse phase.
It means that star formation is a heterogeneous process \citep{Kuffmeier+2017,Bate2018,Lebreuilly+2023}. 

In this paper, we study the role of infall in determining and modifying the angular momentum vector of the star-disk systems ('star-disk spin') and the implications for the orientation and alignment of the disk, as well as the connection to the binarity/multiplicity of the stars \citep{Offner+2023}.
We structure the paper in the conventional way of introduction (section 1), methods (section 2), results (section 3), discussion (section 4) and conclusion (section 5). 

\section{Methods}
\subsection{Recap of the 3D MHD simulation} 
The analysis presented in this paper is based on the simulation data used in \cite{Kuffmeier+2023,Jensen+2023,Jorgensen+2022}, and the simulation design and conceptual setup was presented in depth in \cite{Haugbolle+2018}. 
We therefore only briefly recap the major features of the model and refer the reader to the aforementioned papers for more details.

The 3D simulations were carried out with the ideal magnetohydrodynamical (MHD) version of the adaptive mesh refinement (AMR) code \ramses. 
The simulations have a minimum level of refinement of $l_{\rm ref}=9$ with respect to the total length of the box, which leads to a root grid of $(2^9)^3=512^3$ cells, such that even the least resolved cells do not exceed $\approx 1600$ au. 
The maximum level of refinement with respect to the total length of the box is 15, which implies a resolution of $l_{\rm box}/2^{l_{\rm ref}}=(4 \unit{pc})/2^{15}\approx25$ au for cells at the highest resolution. 
Cells are selected for refinement if they exceed level dependent threshold densities keeping the number of cells per Jeans length larger than 28.8 cells except at the highest level of refinement, where sink particles are introduced.  In this way the densest regions in which the stars form are resolved at the highest resolution.
Moreover, \ramses\ use a graded octree structure and neighboring cells are only allowed to differ by at most one level, which secures gradual refinement towards regions of higher densities and at shock fronts.

The domain is a cubical box with a box length of 4 pc and we apply periodic boundary conditions. 
The initial mass of the box is 3000 M$_{\odot}$ and the initial magnetization is $B=7.2 \mu$G, which implies an Alfv{\'e}nic Mach number of $\mathcal{M}_{\rm A}\equiv \sigma_{\rm v}/v_{\rm A}=$5, where $\mathcal{M}_{\rm A}$ is the average Alfv{\'e}n speed $v_{\rm A} = \frac{B}{\sqrt{4 \pi \rho}}$ with average density $\rho$. 
The gas is set to be isothermal at a temperature of 10 K. 
We drive the turbulence in the box through random solenoidal accelerations with power in an inertial range of wavenumbers $1\le k \le 2$.
The amplitude of the driving is such that the sonic Mach number $\mathcal{M}_{\rm s}\equiv \sigma_{\rm v}/c_{\rm s}$, where $\sigma_{\rm v}$ is the three-dimensional rms velocity and $c_{\rm s}$ is the isothermal speed of sound, is on average $\mathcal{M}_{\rm s}=10$.
The virial parameter $\alpha_{\rm vir}= 5 \sigma_{\rm v}^2 R/ (3 G M_{\rm box} )$, where $G$ is the gravitational constant and $R=L_{\rm box}/2$ is the characteristic dynamical length scale in the box, is  0.83. First the box is evolved for 20 dynamical time-scales to erase any trace of the initial conditions, and then self-gravity and a recipe for star formation is introduced. 
Compared to observations, the conditions are most similar to the star-forming region of Perseus \citep[see observations by][]{Arce+2010} as discussed in \cite{KuruwitaHaugboelle2022}.

The \ramses\ version used for the underlying models is customized with respect to the public version. 
It includes a recipe for sink particles and tracer particles. 
Sink particles act as gravitational sources and record accretion. 
They are used as substitutes for protostars to avoid catastrophic refinement, while still accounting for the formation of stars.
The creation and accretion recipes of the sinks are described in detail in section 2 of \cite{Haugbolle+2018}, and the distribution of the stellar mass function at the end of the simulation is in very good agreement with observations.

To follow the trajectories of the gas in the domain, we use (Lagrangian) tracer particles. 
The tracer particles are passively advected with the flow of the gas in the domain. They do not cause any back-reaction on the gas dynamics. 
We inject 1 tracer particle per root grid cell (corresponding to 134 217 728 tracer particles) after 20 dynamical timescales, at the point of the simulation when gravity is turned on. The underlying cell mass at the root grid level determines the mass of each particle. 
The relevant time span for the analysis presented in this paper is about 2 million years from the formation of the first sink particle in the simulation until the last snapshot of the simulation. The time step between the individual output files is $\Delta t= 5$ kyr.

\subsection{Measuring the spin of star-disk systems with tracer particles} 
We use the tracer particles that accreted onto the individual sinks as a proxy for the orientation of the star-disk system.\footnote{A recipe to track the stellar spin following the method presented in \cite[][]{Federrath+2014} was recently implemented \citep{Dalsgaard+2024}. 
However, the high-resolution run (root-grid of $512^3$ cells) that was used for the analysis in this paper was carried out before the recipe was implemented.}
Each tracer particle gets a flag as soon as it accretes onto a sink particle. 
The time interval between the outputs was $\Delta t = 5$ kyr.
It implies that we measure the time of accretion of a tracer particle as $t_{\rm acc}= \delta t_{\rm create} + n \Delta t$, where $\delta t_{\rm create}$ is the sink age at the first output (the sink data is stored with much higher time frequency such that $\delta t_{\rm create}<5$ kyr) and $n$ is a non-negative integer corresponding to the output after sink formation starting from 0 when the tracer particle is marked as `accreted'.
The spin is based on the sum of the angular momentum of each accreting tracer particle at the snapshot \textit{before} the tracer particle accreted. 
The spin of sink $j$ is based on the angular momentum of $N_{\rm acc}(j)$ tracer particles that are within 0 to 5 kyr prior to accretion: 
\begin{equation}
\mathbf{L}_{j} = \sum_i^{N_{\rm acc}(j)} m_{i}(t_{\rm acc}-\Delta t) \ \mathbf{r}_i(t_{\rm acc}-\Delta t) \times \mathbf{v}_i(t_{\rm acc}-\Delta t) 
\end{equation}

The focus of this work is to investigate the role of infall on the \textit{orientation} of the rotation axis of star-disk systems.
Therefore, using the orientation based on the accreting tracer particles as a proxy for the orientation of the angular momentum vector of star-disk systems is an appropriate assumption for our purpose.

\section{Results}
Computing the angular momentum of the accreting tracer particle for each star allows us to investigate the accumulation of angular momentum. 
As the disks are not resolved in the simulations, we refer to the derived sum of the accreting tracer particles per sink as the star-disk spin. 
We systematically overestimate the magnitude of the accumulated angular momentum because we are missing possible transport processes on the disk scales such as magnetic braking and corresponding transport of angular momentum out of the disk by magnetic fields and outflows.
Therefore, we refrain from including it in our analysis.
However, the orientation of the derived spin vectors provides us with valuable information about the properties and effects of infall.
In the following, we present results of our analysis in which we analyzed the evolution of the spin orientation for stars of various masses over a time span of $1.2$ Myr, measuring the anisotropy of the accretion process and how the accretion process relates to multiplicity.

\subsection{Evolution of star-disk orientation}
\begin{figure}
\includegraphics[width=\linewidth]{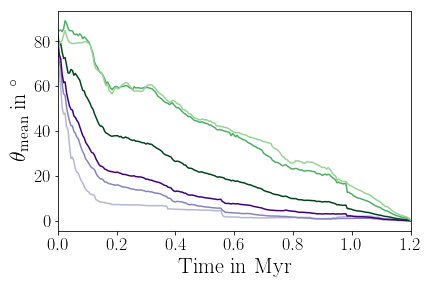}
\includegraphics[width=\linewidth]{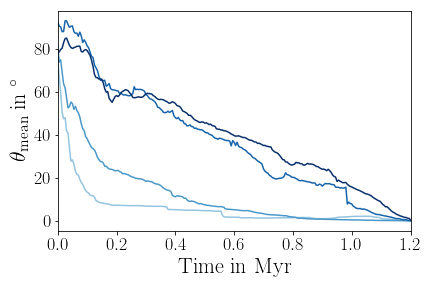}
\caption{Evolution of average angle $\theta$ of star-disk spin with respect to the spin after $t=1.2$ Myr (upper panel) and $t=0.5$ Myr (lower panel) for various upper/lower limits of stellar masses.
The average angle is computed based on stars that have an age of more than 1.2 Myr at the end of the simulation.
Upper panel: The light to dark purple lines mark the mean average relative angle of orientation with respect to the orientation at $t=1.2$ Myr for stars with masses $M_{*}$(1.2 Myr) \textit{less} than $0.2 $M$_{\odot}$, $1 $M$_{\odot}$ and $2 $M$_{\odot}$. Dark to light green lines mark the mean average angle of orientation with respect to the orientation at $t=1.2$ Myr for stars with final masses \textit{greater} than $0.2 $M$_{\odot}$, $1 $M$_{\odot}$ and $2 $M$_{\odot}$.
Lower panel: The light to dark blue lines mark the average
mean angle of orientation with respect to the orientation at $t=1.2$ Myr for stars with masses in the mass ranges of $M_{*}$(1.2 Myr)$<0.2 $M$_{\odot}$, 
0.2 M$_{\odot}<M_{*}$(1.2 Myr)$<1 $M$_{\odot}$,
1 M$_{\odot}<M_{*}$(1.2 Myr)$<2 $M$_{\odot}$, 
and
M$_{*}$(1.2 Myr)$>$2 M$_{\odot}$}.
The upper panel can be directly compared with the results presented in \cite{Kuffmeier+2023}, while the evolution for mass ranges shown in lower panel is arguably more helpful for comparison with observations.  
\label{fig:t_thetamean_1p2M}
\end{figure}

\begin{figure}
\includegraphics[width=\linewidth]{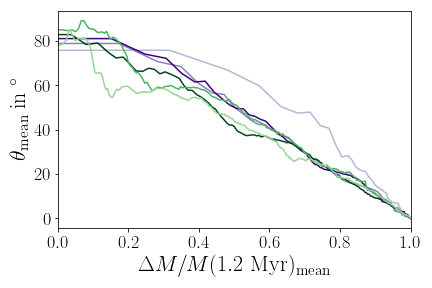}
\includegraphics[width=\linewidth]{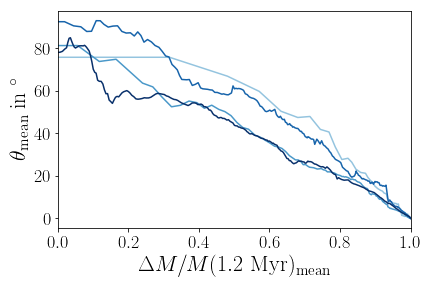}
\caption{Same as \Fig{t_thetamean_1p2M}, but for the mean of $\theta$ as a function of mean of relative mass accrual instead of time.}
\label{fig:dMmean_thetamean_1p2M}
\end{figure}

To compare the spin evolution over the same extended time interval, we analyze the spin accrual of the 104 stars that evolved for at least 1.2 Myr in the simulation.
The choice of 1.2 Myr is made to allow a sample of more than 100 stars, while also accounting for infall beyond 1 Myr in the simulation as much as possible.
We show the evolution of the average mean in \Fig{t_thetamean_1p2M}) (and median in \Fig{t_thetamedian_1p2M}) angle of the star-disk spin with respect to the spin orientation at $1.2$ Myr of the selected 104 stars. 
We find a clear overall dependency of the spin evolution on the stellar mass $M_*$ at $t=1.2$ Myr. 
On average, the higher $M_*(t=1.2 \unit{Myr})$, the longer it takes for the stars to obtain the reference spin orientation at $t=1.2$ Myr. 
This result can be understood in the context of the average mass accretion histories of the stars reported in \citet{Kuffmeier+2023}. 
In that paper, we found a two-phase accretion process of stars consisting of an initial collapse phase followed by a post-collapse phase.
During the post-collapse phase, the star is fed by additional material that is initially not bound to the collapsing progenitor core. 
The higher the final mass of the star, the larger is the contribution of the post-collapse infall phase.
Considering such a process, lower mass stars tend to reach their final spin orientation more quickly than higher mass stars that undergo a prolonged formation phase fueled by infall of material misaligned with respect to the stellar orientation at this point.
We also find a kink or a change in the profile of the mean angles within 100 kyr to 200 kyr for all mass bins.
We interpret the kink as the transition from the collapse to the post-collapse infall phase, identified in the two-phase accretion scenario.

In the idealized star formation picture where the star is solely fed by material from an isolated core with uniform rotation, it would be possible to have a prolonged formation time for higher mass stars in the case of sufficiently large cores and under the assumption without fragmentation.
However, the star would not show deviations from its rotational axis when it is solely fed with material that has the same rotational axis.
Interestingly, the lower mass stars that on average form more rapidly and from a mass reservoir of smaller radial extent also have initially different spins compared to their spin at $t=1.2$ Myr. 

This effect can be better appreciated in \Fig{dMmean_thetamean_1p2M}, where we plot the evolution of the spin angle with respect to the spin orientation at $t=1.2$ Myr over relative mass accrual instead of time. 
In contrast to the evolution over time, the plot does not show a clear sign of mass dependency in the spin accrual process in terms of relative mass gain.
On average and independent of the mass, the initial spin axis is initially misaligned by about $80^{\circ}$ to $90^{\circ}$ compared to the spin axis at $1.2$ Myr.
The stars approach their final spin by an approximate rate of about $8^{\circ}$ to $9^{\circ}$ per $10 \%$ stellar mass accrual. 
This shows the role of turbulence in the accretion process already during the early collapse stages \citep{Seifried+2012}.
When the infalling material stems from a turbulent medium, the accretion process is more chaotic in nature, and hence variations in the orientation of the angular momentum vector are expected. 
Our analysis demonstrates that this is the case in agreement with previous results by \cite{Fielding+2015}.

\begin{figure}
\includegraphics[width=\linewidth]{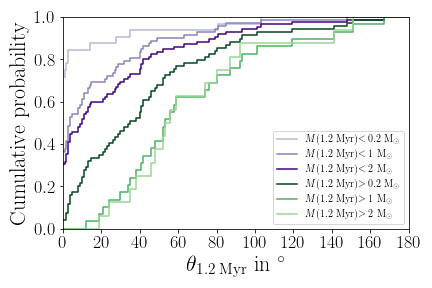}
\includegraphics[width=\linewidth]{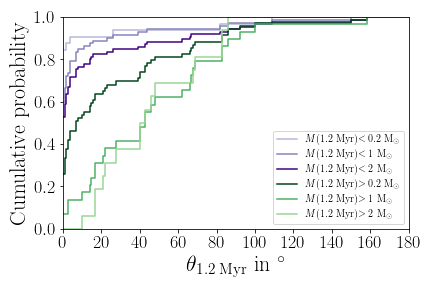}
    \caption{Cumulative probability distribution of changes in stellar spin by angle $\theta$ with respect to spin at reference time $t=200$ kyr (upper panel) and $t=500$ kyr (lower panel). The line colors indicate the same stellar mass thresholds as in the top panel of \Fig{t_thetamean_1p2M}. 
}\label{fig:thetahist_1p2M}

\end{figure}

We emphasize that the reported dependency applies to the \textit{average} of the considered mass distributions sample. 
The distribution of the spin angles at $t=200$ kyr (upper panel) and $t=500$ kyr (lower panel) in \Fig{thetahist_1p2M} shows this. 
For instance, at $t=200$ kyr, for stars with $M_*(1.2 \unit{Myr})>1$ M$_{\odot}$, about 60 \% have a spin that is $60^{\circ}$ or less from the spin orientation at $M_*(1.2 \unit{Myr})$. 
The remaining $40 \%$ of the stars with $M_*(1.2 \unit{Myr})>1$ M$_{\odot}$, have spins that range from angles of $60^{\circ}$ up to almost $180^{\circ}$. 
In fact, almost $20 \%$ of the stars with $M_*(1.2 \unit{Myr})>1$ M$_{\odot}$ have spins at 200 kyr that differ by more than $90^{\circ}$ in angle with respect to their spin at $t=1.2$ Myr. 
This implies that there is enough retrograde infall that the entire system readjusts into a configuration where the entire system changes its orientation by more than $90^{\circ}$.
We note that these drastic changes apply to the spin of the entire star-disk system. 
Considering that the disk is typically only $\sim1 \%$ of the stellar mass, we expect the effect of misaligned infall on the disk alone to be even stronger than shown for the proxy of star-disk spin.

Models of infalling gas with different angular momentum vectors onto an existing disk showed that such infall can lead to a configuration of misaligned inner and outer disks \citep{Thies+2011,Kuffmeier+2021}. 
The results presented above confirm that assuming initial conditions of misaligned infall is justified.
In fact, misaligned infall frequently occurs in this simulation especially for stars with masses beyond a few $0.1$ M$_{\odot}$.

\subsection{Anisotropic accretion and multiplicity}
\begin{figure}
\includegraphics[width=\linewidth]{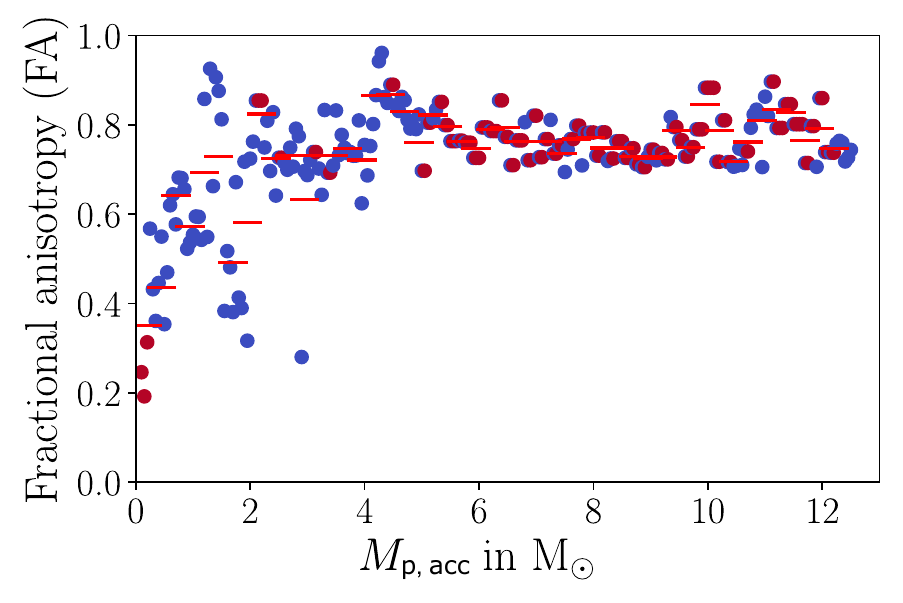}
\caption{Fractional anisotropy (FA) of the accreting material for a star in the simulation. Each dot represents FA for accrual of a mass bin of $0.05$ M$_{\odot}$. Blue colors mean that the star was part of a multiple system while accreting the corresponding mass, red dots mean that it accreted its material as a single star. The red horizontal lines represent the average of FA over five mass bins, i.e., over $0.25$ M$_{\odot}$.  
}\label{fig:dm_FA_89}
\end{figure}

As shown in numerous studies, protostellar accretion happens along accretion channels and sheets \citep[e.g.,][]{Offner+2010,Seifried+2012,Seifried+2013,Santos-Lima+2012,Joos+2013,Li+2014,Kuffmeier+2017,Heigl+2024}.
In other words, accretion is anisotropic rather than spherically isotropic.
This conclusion generally relies on filamentary accretion shown in the form of 3D visualizations or projection maps that visualize infall-outflow motions at a given radius and there is general consensus about filamentary accretion in the star formation community nowadays. 
While the reported results are intriguing and informative, we are, however, lacking a quantitative measure to state the level of (an)isotropy in the accretion process. 

In this paper, we overcome this by using a quantity that is usually used to measure the (an)isotropy in the context of diffusion processes, namely fractional anisotropy ($\mathrm{FA}$). 
It determines whether diffusion occurs rather along a predominant direction (anisotropic case) or whether it occurs unrestricted along all directions (isotropic case), and it is defined as 
\begin{equation}
\mathrm{FA} = \sqrt{ \frac{1}{2}} \frac{\sqrt{(\lambda_1 - \lambda_2)^2 + (\lambda_2 - \lambda_3)^2 + (\lambda_3 - \lambda_1)^2}}{\sqrt{\lambda_1^2 + \lambda_2^2 + \lambda_3^2}},
\end{equation}
with $\lambda_1$, $\lambda_2$ and  $\lambda_3$ being the eigenvalues of the covariance matrix corresponding to the points in the system.  
$\mathrm{FA}=0$ corresponds to a perfectly isotropic case, $\mathrm{FA}=1$ means maximally anisotropic.
In our case, the points for the covariance matrix are determined by the components of the angular momentum vector of the accreting tracer particles.

We compute $\rm FA$ per accreting mass interval of $0.05$ M$_{\odot}$ for each star. As an example, 
\Fig{dm_FA_89} shows the evolution of FA for a massive star in the simulation. 

Apart from the FA of each mass bin, \Fig{dm_FA_89} also shows whether the star accreted the corresponding mass fraction as a single star or as part of a multiple system. 
To determine whether the star accretes as a single or multiple star, we analyze the full multiplicity distribution for each snapshot, following the recipe of  \citet{KuruwitaHaugboelle2022} to find all bound systems within a critical distance that we set $2 \times 10^4$ au. 
If the selected star is part of a bound systems within $2 \times 10^4$ au during the accrual of $\Delta m_{\rm p}=0.05 \unit{M}_{\odot}$ at any snapshot, we mark it as accreting as part of a multiple system during that interval (dark blue dots). 
If it misses a bound member within $2 \times 10^4$ au, we mark it as accreting as a single star during that interval (red dots). 
The plot shows that this star primarily accretes the first half of its mass with other stars in its vicinity, while in the second half it accretes more as a single star.
Using the information obtained for all of the selected stars, we compare the accretion histories of the different stars.
This allows us to measure the (an)isotropy of the accreting material for all of the stars (see section \ref{sec_stats}), to 
analyze whether there is an evolutionary trend in the accretion mode (see section \ref{sec_early-late}), and investigate the influence of multiplicity on the accretion process (see section \ref{role-of-multi}). 

\subsubsection{(An)isotropy of accretion for all stars}
\label{sec_stats}
\begin{figure}
\includegraphics[width=\linewidth]{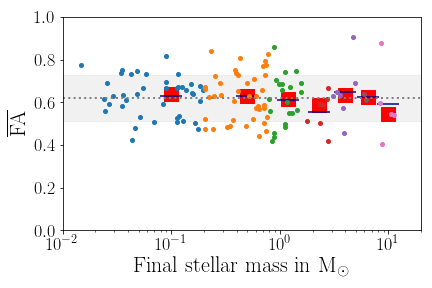}
\caption{$\rm FA$ of the first $\Delta m_{\rm p} = 0.05 \unit{M}_{\odot}$ with respect to the average $\rm FA$ of the accreting material over the mass at the end of the simulation. 
The dots show $\frac{\rm FA_{0}}{\overline{\rm FA}}$ of all the 124 stars that have reached an age of at least 974 kyr. 
Note that the efficiency factor of sink accretion is set to $50\%$ to account for the mass loss via outflow. That means the stellar mass only increases by 0.025 M$_{\odot}$ for accretion of 0.05 M$_{\odot}$.
Light blue corresponds to 41 stars with $M_*<0.2 \unit{M}_{\odot}$, orange to 41 stars in the range $0.2 \unit{M}_{\odot} \le M_*<0.8 \unit{M}_{\odot}$, green to 21 stars in the range $0.8 \unit{M}_{\odot} \le M_*<1.6 \unit{M}_{\odot}$, dark-red to 7 stars in the range $1.6 \unit{M}_{\odot} \le M_*<3 \unit{M}_{\odot}$, purple to 7 stars in the range $3 \unit{M}_{\odot} \le M_*< 5 \unit{M}_{\odot}$, brown to 2 stars in the range $5 \unit{M}_{\odot} \le M_*<8 \unit{M}_{\odot}$ and violet to 5 stars with $ M_* > 8 \unit{M}_{\odot}$.
The red squares correspond to the median per bin, the dark-blue horizontal lines show the mean per bin. 
The grey dotted line shows the mean value of all stars and the light-grey area marks the corresponding standard deviation.  
}\label{fig:Mstar_FA}

\end{figure}

We measure the fractional anisotropy of the accreting material for all 124 stars that evolve to at least $974$ kyr over mass accrual of $\Delta m_{\rm p} = 0.05$ M$_{\odot}$ as shown for the exemplary star shown in \Fig{dm_FA_89}. 
The lower age limit is used in order to include the reference star that was highlighted as a prime example of late accretion in \cite{Kuffmeier+2023}. 
In \Fig{Mstar_FA}, we plot the average FA for the 124 stars over the final mass of the corresponding stars. 
Overall, we find an average of $\mathrm{FA} = 0.62 \pm 0.11$ over all stars, which shows that infall generally deviates from isotropic collapse. 
This corroborates the changes of the orientation of the star-disk spin vector described above. 
At first sight, the average FA alone does not reveal a striking dependency on stellar mass though the scatter is slightly larger for increasing final mass.  
We interpret these results as a sign of turbulence generally implying filamentary accretion regardless of the accretion phase, and the scatter as a sign of a heterogeneous accretion process in molecular clouds. 

\subsubsection{Variations between early and late accretion phase}
\label{sec_early-late}
\begin{figure}
\includegraphics[width=\linewidth]{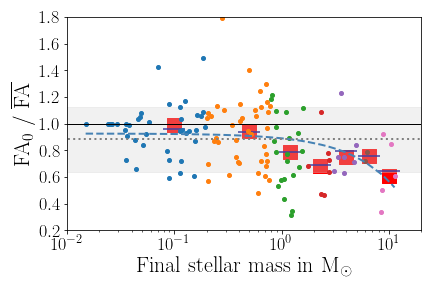}
\caption{$\rm FA$ of the first $0.05 \unit{M}_{\odot}$ with respect to the average $\rm FA$ of the accreting material over the stellar mass at the end of the simulation. 
The dots show $\frac{\rm FA_{0}}{\overline{\rm FA}}$ of all the 124 stars that have reached an age of at least 974 kyr. Light blue corresponds to 41 stars with $M_*<0.2 \unit{M}_{\odot}$, orange to 41 stars in the range $0.2 \unit{M}_{\odot} \le M_*<0.8 \unit{M}_{\odot}$, green to 21 stars in the range $0.8 \unit{M}_{\odot} \le M_*<1.6 \unit{M}_{\odot}$, dark-red to 7 stars in the range $1.6 \unit{M}_{\odot} \le M_*<3 \unit{M}_{\odot}$, purple to 7 stars in the range $3 \unit{M}_{\odot} \le M_*< 5 \unit{M}_{\odot}$, brown to 2 stars in the range $5 \unit{M}_{\odot} \le M_*<8 \unit{M}_{\odot}$ and violet to 5 stars with $ M_* > 8 \unit{M}_{\odot}$.
The red squares correspond to the median per bin, the dark-blue horizontal lines show the mean per bin. 
The grey dotted line shows the mean value of all stars and the light-grey line is the corresponding standard deviation.  
The light-blue dashed line shows the polynomial fit based on all points using numpy's polyfit function.
To distinguish between initially more/less isotropy, we display a black solid line at value 1. 
}\label{fig:Mstar_FAratio}

\end{figure}

As highlighted in \cite{Kuffmeier+2023}, star formation progresses in two stages \citep[arguably with a third stage if accounting for the mass accrual prior to gravitational collapse of the core as part of the process, see][]{Padoan+2020}. 
The first phase is conceptually similar to the classical picture of a collapsing core, while the second phase consists of a sweep-up of material when the star evolves in the Giant Molecular Cloud. 
We consider the second phase similar to Bondi-Hoyle accretion \citep[][]{Scicluna+2014} but with an impact parameter of the encountering inflow of material \citep{Dullemond+2019,Kuffmeier+2020}.
This mode of star formation is also in agreement with observations of accretion streamers that feed star-disk systems with fresh material from the molecular cloud environment \citep{Pineda+2019,Alves+2020,Huang+2020,Huang+2021,Huang+2023,Valdivia-Mena+2022,Valdivia-Mena+2023,Gupta+2023,Cacciapuoti+2024}.
In particular, a recent analysis of streamer candidates in NGC 1333 by \cite{Valdivia-Mena+2024} revealed a fraction of about 40 $\%$.

Against this background, we expect the accretion process to be more isotropic in the beginning when the star forms from a local mass reservoir, while the post-collapse phase is more anisotropic. 
If the initial phase is more isotropic than the late accretion history, we expect that $\rm FA$ at early times is smaller compared to the average $\rm FA$.
Moreover, the contribution of post-collapse infall, on average, increases with increasing final stellar mass. 
Therefore, we expect this correlation to be more enhanced for higher mass stars that experience a larger contribution of post-collapse infall than lower mass stars that experience only little post-collapse infall. 

The evolution of FA for the exemplary star shown in \Fig{dm_FA_89} plot shows such a trend from more isotropic accretion with FA as low as 0.2 for the first accreting 0.1 M$_{\odot}$ at the beginning to more anisotropic accretion at later stages with FA between 0.7 and about 0.9. 
To investigate this trend in more detail,
we compute the ratio of fractional anisotropy for the first $0.05 \unit{M}_{\odot}$ of accreting mass and the mean of $\rm FA$ given by 
\begin{equation}
    \frac{\rm FA_{0}}{\overline{\rm FA}} =     \frac{\rm FA(0<M_{\rm p, acc}<0.05 \unit{M}_{\odot})}{\overline{\rm FA}}
\end{equation}
for each star.
Note that we apply an accretion efficiency factor of $50 \%$, such that only $50 \%$ of the cell mass that meets the criteria of accretion is added to the sink mass. 
This efficiency factor is used to account for mass loss via outflows.

The results are shown in \Fig{Mstar_FAratio}. 
We find that the mean ratio over all 124 stars is $\approx 0.88\pm0.25$, hence below 1 as expected. 
Applying numpy's polyfit function to the data points, we fit the data with a straight line with slope $\approx -0.036$ and offset $\approx 0.927$.
The decreasing ratio is a consequence of the larger relative contribution of late infall to the final mass for higher-mass stars.
Despite a significant scatter, this analysis confirms that late infall is, on average, more anisotropic than the early collapse phase.

\subsubsection{The role of multiplicity in the accrual of star-disk spin}
\label{role-of-multi}
\begin{figure}
\includegraphics[width=\linewidth]{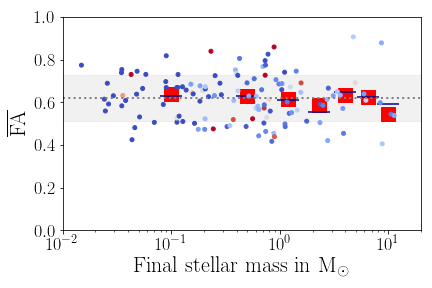}
\includegraphics[width=\linewidth]{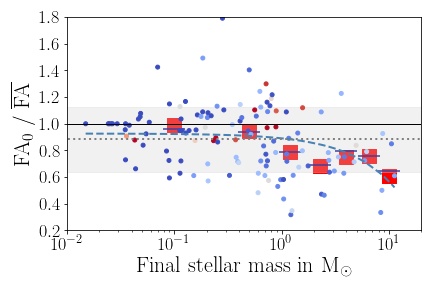}
\caption{
Top/bottom panel: same as \Fig{Mstar_FA}/\Fig{Mstar_FAratio} except that dots of individual stars are colored according to their evolution as a single star or as part of a multiple system. 
The redder (bluer) the dot, the higher is the fraction of time that the star evolved as a single star (multiple system) while accreting. 
}\label{fig:Mstar_FA_multi}

\end{figure}

\begin{figure}
\includegraphics[width=\linewidth]{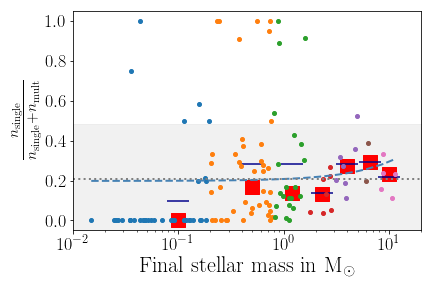}
\includegraphics[width=\linewidth]{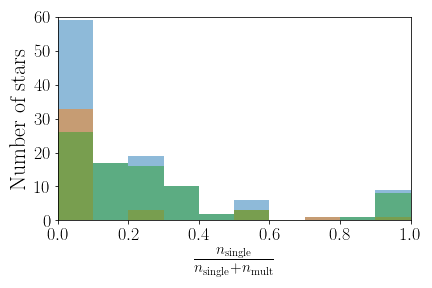}
\caption{Top panel:
Fraction of output snapshots when individual stars accrete mass as isolated single stars without any neighboring star within a radial distance of $2\times 10^4$ au and total number of output snapshots when accretion is ongoing $\frac{n_{\rm single}}{n_{\rm single}+n_{\rm mult}}$ over final stellar mass.
Each filled circle represents one of the 124 stars that have reached an age of at least 974 kyr. 
Light blue corresponds to 41 stars with $M_*<0.2 \unit{M}_{\odot}$, orange to 41 stars in the range $0.2 \unit{M}_{\odot} \le M_*<0.8 \unit{M}_{\odot}$, green to 21 stars in the range $0.8 \unit{M}_{\odot} \le M_*<1.6 \unit{M}_{\odot}$, dark-red to 7 stars in the range $1.6 \unit{M}_{\odot} \le M_*<3 \unit{M}_{\odot}$, purple to 7 stars in the range $3 \unit{M}_{\odot} \le M_*< 5 \unit{M}_{\odot}$, brown to 2 stars in the range $5 \unit{M}_{\odot} \le M_*<8 \unit{M}_{\odot}$ and violet to 5 stars with $ M_* > 8 \unit{M}_{\odot}$.
The red squares correspond to the median per bin, the dark-blue horizontal lines show the mean per bin. 
The grey dotted line shows the mean value of all stars and the light-grey line is the corresponding standard deviation.  
The light-blue dashed line shows the line fit based on all points using numpy's polyfit function.
Bottom panel:
Histogram of $\frac{n_{\rm single}}{n_{\rm single}+n_{\rm mult}}$ for all 124 stars that have reached at least 974 kyr of age (light blue bars), 41 stars with final stellar mass $M_{*, \rm final}<0.2$ M$_{\odot}$ (brown-orange bars) and 83 stars with $M_{*, \rm final}>0.2$ M$_{\odot}$ (green bars). 
}\label{fig:Mstar_s-to-m}

\end{figure}

Considering the mass fraction that stars accrete as a single or as part of multiple stars (\Fig{Mstar_FA_multi}), we do not find an immediately striking correlation of FA with multiplicity. 
For instance, the plot shows two stars with almost identical final masses of about $0.2$ M$_{\odot}$ almost exclusively accreting their mass as single stars, but very different fractional anisotropies of FA$> 0.8$ and FA$< 0.5$. 
Similarly, two stars with final mass of about 0.8 M$_{\odot}$ that mostly accrete as part of multiple systems have values of FA $\approx 0.4$ and FA$> 0.8$.  
Nevertheless, there are a few indications for mass-dependent differences in the accretion process. 
Considering to which degree each star accreted its material as part of a multiple system or as a single star, the lower panel of \Fig{Mstar_FAratio} hints at a slightly higher ratio $\frac{\rm FA_{0}}{\overline{\rm FA}}$ for stars that mostly accrete in isolation. 
Given the relatively low number of about a dozen stars that are categorized as primarily accreting as single stars and the corresponding scatter, the evidence is strong, but it suggests that isolated stars are more likely to occur and accrete in a less turbulent medium compared to stars associated with clusters. 

The upper panel in \Fig{Mstar_FA_multi} suggests that lower mass stars more often accrete (almost) all of their mass while being classified as part of a multiple system, while higher mass stars tend to accrete part of their mass as single stars in agreement with the result shown for the example star in \Fig{dm_FA_89}. 
This correlation is more visible when plotting the fraction of mass that each star accretes as a single star, $\frac{n_{\rm single}}{n_{\rm single}+n_{\rm mult}}$ (see top panel in \Fig{Mstar_s-to-m}). 
Fitting the data with a line, we find a very mild positive correlation with mass.
The slope is $0.01$ and the offset is $0.20$.
We point out, however, that the average ratio of all stars is $\overline{\frac{n_{\rm single}}{n_{\rm single}+n_{\rm mult}}} = 0.21 \pm 0.27$, which shows that there is significant scatter in the data. 
The scatter appears to be smaller for higher mass stars of final mass above $\approx 1$ M$_{\odot}$.
This is reflecting that the lower the final mass of the star, the quicker they accrete their mass and the lower the number of mass bins $\Delta m_{\rm p}=0.05$ M$_{\odot}$ that are considered. 

Regardless of the uncertainties and the scatter, the majority of stars accrete their mass with other stars in their vicinity. 
The histogram of $\frac{n_{\rm single}}{n_{\rm single}+n_{\rm mult}}$ (bottom panel in \Fig{Mstar_s-to-m}) shows that only 17 of the 124 stars ($\approx13.7 \%$) accrete more than half of their mass while being classified as an isolated single star. 
For stars with final masses less than $0.2$ M$_{\odot}$, the fraction is yet smaller ($5/41\approx 12.2 \%$), while for stars with masses $M_*>0.2$ M$_{\odot}$, the fraction is slightly higher ($12/83\approx 14.5 \%$).

\subsection{Spin in stellar multiples}
Using the multiplicity analysis with a cut-off distance of $2 \times 10^4$ au and allowing for a maximum of 10 stars per system in the search, we find that several stellar systems of order binary or higher have formed by the end of the simulation. 
In particular, we find in total 24 systems that consist of three or more stars at time $t=1.91$ Myr. 
Using the measure of FA, we analyze the anisotropy in orientation of the stellar spins in these systems. 
Given the small number of stars in the protostellar multiples and the relatively low number of multiple systems, however, we point out that the following results derived from the analysis of this sample are less robust.
They should be understood as indicative and as early constraints for future analyses that are based on larger sample sizes as discussed in section \ref{discussion}.

\begin{figure}
\includegraphics[width=\linewidth]{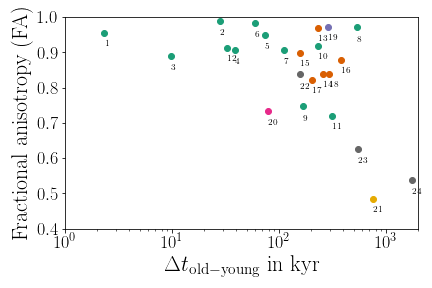}
\includegraphics[width=\linewidth]{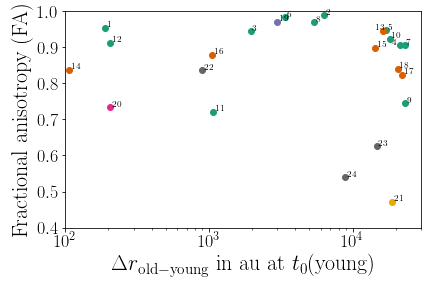}
\includegraphics[width=\linewidth]{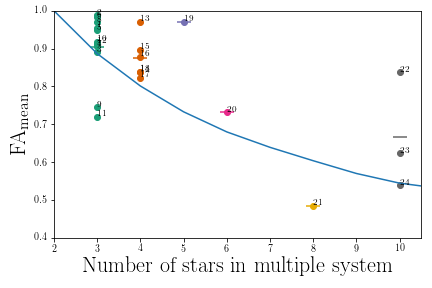}
\caption{Fractional anisotropy of multiple systems at $t\approx1.91$ Myr after formation of the first star in the simulation.
Top panel: fractional anisotropy ($\rm FA$) versus difference between formation time of oldest and youngest member. 
Middle panel: $\rm FA$ versus distance between the oldest and youngest member at the formation time of the youngest member. 
Bottom panel: $\rm FA$ versus number of stars in the corresponding system. The points show the values of the systems and the horizontal bars show the averages for systems with the same number stars. 
The blue line shows the expectation value for an entirely random distribution computed with 10000 iterations per number of stars using the random sample routine of numpy \citep{Harris2020array}. 
At this point in time, there are 12 triple systems (green points), 6 systems with 4 stars (orange points), 1 system with 5 stars (blue-purple point), 1 system with 6 stars (violet points), 1 system with 8 stars (yellow point) and 3 systems with 10 stars (gray points). 
}\label{fig:dt_FA}

\end{figure}

We plot the fractional anisotropy $\rm FA$ of these systems over the age difference between the oldest and youngest star in the system $\Delta t_{\rm old - young}$ (\Fig{dt_FA}).
The plot shows two major properties. 
First, systems with more components tend to have lower values of $\rm FA$ and thereby a more isotropic distribution.
Second, systems in which the age difference between the cluster members are larger have lower values of $\rm FA$ than systems consisting of stars that formed within a shorter amount of time.

We hypothesize that this as a consequence of the formation history of the systems. 
Neighboring stars that form within a few $10^4$ kyr must have formed relatively close to each other likely through core or filament fragmentation in the same region. 
Therefore, they generally gain their mass and angular momentum from the same reservoir that has the same net angular momentum. 
Neighboring stars that have age differences of $10^2$ kyr or more, however, 
formed in different regions and only became neighbors at later stages via dynamical capture. 
That implies that the stars gained a significant part of their mass and angular momentum from different reservoirs.
It is likely that the net angular momentum vectors of these regions pointed in different directions, and hence the stellar spins in systems that originate from dynamical capture are more likely to be misaligned \citep[e.g.,][]{Pineda+2012,Brinch+2016,Maureira+2020,Jorgensen+2022}.
That explains the more isotropic distribution of the spins in these systems compared to systems that formed from the same mass reservoir via core collapse and core fragmentation.
To test this hypothesis, we compared the distance between the oldest and youngest member at the time of formation of the youngest member for the systems $r_{\rm old-young}(t=t_0(\rm young))$ displayed in the top panel of \Fig{dt_FA}. 
If the hypothesis is true, we expect a larger distance for the systems formed via dynamical capture than for those that formed via fragmentation in the same region.
In the middle panel of \Fig{dt_FA}, we plot the measured $\rm FA$ shown in \Fig{dt_FA}, but this time over $r_{\rm old-young}(t=t_0(\rm young))$. 
Keeping in mind the caveat of the very low sample size as shown in the bottom panel of \Fig{dt_FA},
the plot is in agreement with the expected correlation from the hypothesis that systems forming via dynamical capture tend to have a larger isotropy.

\section{Discussion}
\label{discussion}
\subsection{Limitations of measuring the (an)isotropy of individual stars}
The results of stellar spin evolution presented in this paper are obtained by the analysis of accreting tracer particles. 
Even though the simulation used in this paper has an unprecedented mass resolution and spatial resolution for a global model of a starforming region and required approximately 100 million core-hours to integrate, the smallest cell size is still 25 au. The downside of this is that we do not resolve accretion disks.
In reality, a disk forms and the accretion process likely becomes more anisotropic if the star is predominantly fed from the disk.
We also only have a record of tracer particles with a cadence of $5$ kyr. 
Therefore, the results should be understood as a first constraint on how infall can modify the orientation of disks. 
Earlier models considering infall onto disks already demonstrated that infall may lead to instabilities \citep{Lesur+2015,Bae+2015,Kuffmeier+2018,Kuznetsova+2022}.
Certainly, it will be crucial to account for the presence of the disk self-consistently in future models to obtain a better understanding of the role of infall onto the disk dynamics and the implications for planet formation.   

Nonetheless, given the constraints in spatial and temporal resolution, it is remarkable how much valuable information can already be obtained about the potential of infall in modifying the orientation of star-disk systems. With the current resolution, we are indeed tracing the anisotropy of the infalling material instead of anisotropic features that are induced by the presence of a disk. Therefore, the result that the accretion process during the post-collapse infall phase is more anisotropic than during the initial collapse phase is robust.

\subsection{Limitations of measuring the (an)isotropy of the star-disk spin of protostellar multiples}
There is a serious caveat in the analysis based on FA for stellar systems with relatively few members.
Although the trend of lower $\rm FA$, and hence higher isotropy, for increasing $\Delta t_{\rm old - young}$ is intriguing,
$\rm FA$ in multiple systems of only few members is prone to bias of low number statistics as can be seen by the expectation value of FA for a given number of N points based on randomized distributions shown in \Fig{dt_FA}. 
In the future, we aim for measuring the spin of stars in clusters by carrying out simulations of a larger Giant Molecular Cloud. 
Although the domain and sample size of the underlying simulations analyzed in this paper is too small to derive definite conclusions on the (an)isotropy of the stellar spin in clusters, 
we expect that $\rm FA$ will be a good metric,
when comparing the spin orientation of clusters with more members. 
\Figure{n_FA_mean} shows that the expectation value drops to $<0.2$ for $N>100$ members,
which illustrates the reduced role of small number statistics for larger clusters with a higher number of members. 
We emphasize that the analysis of the anisotropy of the accretion process in section 3.2 is practically unaffected by this bias because the considered mass bins of $0.05$ M$_{\rm odot}$ typically contain $\sim 10^3$ tracer particles.

\subsection{Implications for interpretation of disk observations}
Star formation models that take into account the turbulent dynamics in molecular clouds show that post-collapse infall is a common feature of star formation, especially for stars with final masses beyond a few 0.1 M$_{\odot}$ \citep{Smith+2011,Pelkonen+2021,Kuffmeier+2023}.  
In addition, the analysis presented in this paper demonstrates that infall is anisotropic.  
Such prolonged histories of mass accretion through anisotropic infall are consistent with recent observations of accretion streamers such as Per-emb-2 \citep{Pineda+2019}, Per-emb-50 \citep{Valdivia-Mena+2022}, Barnard 5 \citep{Valdivia-Mena+2023}, L1455 IRS1 \citep{Chou+2016}, IRAS 16544-1604 \citep{Kido+2023}, IRAS 16253-2429 \citep{Aso+2023}, DK Cha \citep{Harada+2023}, GGD 27-MM1 \citep{Fernandez-Lopez+2023},
BHB 1 \citep{Alves+2020}, HL Tau \citep{Yen+2019,Garufi+2021}, M512 \citep{Grant+2021,Cacciapuoti+2024}, AB Aurigae \citep{Grady+1999}, SU Aurigae \citep{Akiyama+2019,Ginski+2021,Labdon+2023}, S Cra \citep{Gupta+2023}, RU Lup \citep{Huang+2020}, GM Aur \citep{Huang+2021} or DR Tau \citep{Huang+2023}.
Moreover, our results show that infall is typically misaligned with respect to the orientation of the star-disk system.  This important result implies that the assumption of misaligned infall in previous models \citep{Thies+2011,Kuffmeier+2021} is justified, which provides strong support for the formation of misaligned disks via late infall traceable as shadows in scattered light observations of disks \citep{Krieger+2024}.
A missing ambitious step in the near future is to combine the two approaches by increasing the resolution of the large-scale models sufficiently to resolve individual infall events and thereby to study how they affect the primordial disk and how often such infall can lead to misaligned disk-systems. This require a new generation ultra-scalable codes that can support the volume and dynamic range needed, such as the exa-scale ready code DISPATCH \citep{Nordlund+2018} being developed in Copenhagen.

\section{Conclusion}
In this paper, we studied the role of infall by analyzing data from a 3D simulation of a molecular cloud of (4 pc)$^3$ in volume that was carried out with the \ramses\ code. 
Inspired by previous results that demonstrated the possibility of mass accretion after the initial collapse phase via post-collapse infall, we study the effect of infall on the orientation of star-disk systems.
Measuring the angular momentum vector of tracer particles at the snapshot prior to accretion onto a star, we find that infall is typically misaligned with respect to the current orientation of the star-disk system. 
As the infall events can contribute a significant mass fraction to the final mass of the star ($\sim 10 \%$), misaligned infall can drastically reorient the rotational axis of the star-disk system.
We also find that, on average, it takes more time for a star-disk system to reach its final spin orientation, the larger its final mass.
This result is consistent and a direct consequence of the mass accretion histories that, on average, the duration of infall lasts longer for increasing final mass of a star. 

Our results show that the orientation of star disk systems is tightly coupled to the amount of material that is infalling. 
In fact, when normalizing the spin orientation of the star-disk systems to the accreting mass relative to the final mass, we find that the spin of the star evolves independently of the final stellar mass from an initial misalignment with respect to its final orientation of $\approx 80^{\circ}$ roughly at a rate of $8^{\circ}$ per $\Delta M=0.1 M_{*, \rm final}$.

We do not resolve the disks in the underlying simulations, but we see significant changes in the orientation during infall for mass infall that contributes ($\sim 10 \%$) to the final mass of the star. 
Unless angular momentum transport is very efficient at the 100 au to 500 au scale, we expect significant reconfiguration of the disk even for infall events of relatively little mass with respect to the final stellar mass ($\lesssim 1 \%$) considering that the disk itself only has a mass of $\sim 1 \%$ of the host star.
We will investigate the effect of infall on the configuration and orientation of the disk further in upcoming zoom-in simulations. 

We also measure to what extent each star in the simulation accreted as a single star or as part of a multiple system. 
The analysis shows that the stars with final masses $\lesssim 0.2$ M$_{\odot}$ are more likely to form and hence accrete their material when they are a member of a system consisting of multiple stars. 
For stars with masses in the range of 0.2 to 1 M$_{\odot}$, the probability is higher to find stars that exclusively accreted their material as a single star. 
However, the majority of stars is part of multiple systems or at least in the vicinity of other stars during most of their accretion period. 
Stars with final stellar masses $\gtrsim 1$ M$_{\odot}$ are often single stars for a small part of their accretion, but are members of multiple systems throughout the majority of their accretion history.

We also provide constraints on the anisotropy of the accretion process. 
To do so, we adapted the measure of fractional anisotropy (FA) that is typically used to give an estimate of the (an)isotropy of a diffusion process. 
We find a mean FA of $\approx 0.6$ averaged over 124 stars that reach an age of $\approx 1$ Myr or more, which means that infall and thereby accretion is anisotropic. 
There is no dependency of the average fractional anisotropy of accreting stars on the final stellar mass.

A comparison of the ratio of FA for the initial $0.05$ M$_{\odot}$ with the mean of FA shows that the accretion process is initially more isotropic compared to later stages.
We also find a trend that the ratio decreases with increasing final stellar mass. 
This is consistent with a mode of accretion, where the star forms from the initial collapse of a dense gas followed by more anisotropic accretion that is funneled by infall through accretion streamers.

\begin{acknowledgements}
The research of MK is supported by H2020 Marie Sk\l{}odowska-Curie Actions (897524) and a Carlsberg Reintegration Fellowship (CF22-1014).
The Tycho HPC facility at the University of Copenhagen, supported by research grants from the Carlsberg, Novo, and Villum foundations, was used for carrying out the simulations, the analysis, and the long-term storage of the results.
\end{acknowledgements}

\bibliographystyle{aa}

\begin{appendix}
\section{Mean and median of spin angles over median and mean masses}
\begin{figure}
\includegraphics[width=\linewidth]{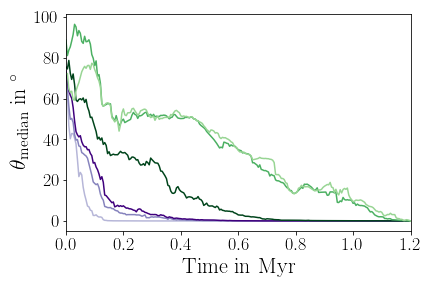}
\includegraphics[width=\linewidth]{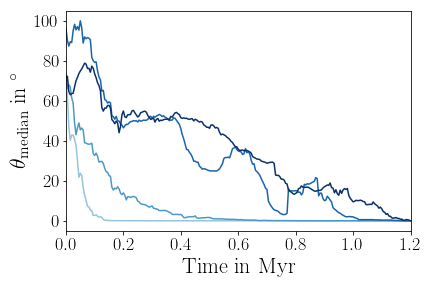}
\caption{Same as \Fig{t_thetamean_1p2M}, but for the median of $\theta$.}
\label{fig:t_thetamedian_1p2M}
\end{figure}

\begin{figure}
\includegraphics[width=\linewidth]{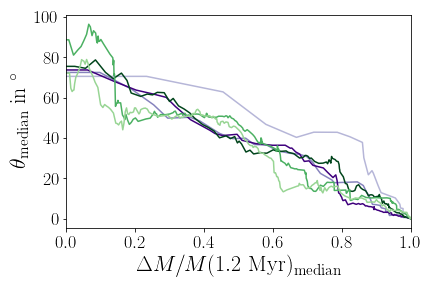}
\includegraphics[width=\linewidth]{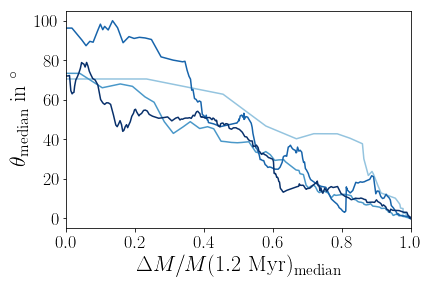}
\caption{Same as \Fig{t_thetamean_1p2M}, but for the median of $\theta$ over median of relative mass accrual.}
\label{fig:dMmedian_thetamedian_1p2M}
\end{figure}

\begin{figure}
\includegraphics[width=\linewidth]{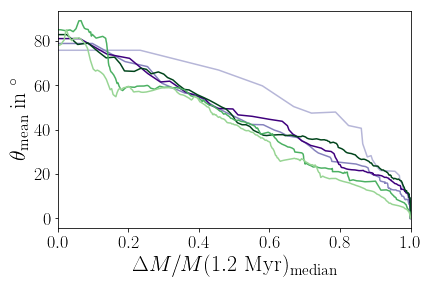}
\includegraphics[width=\linewidth]{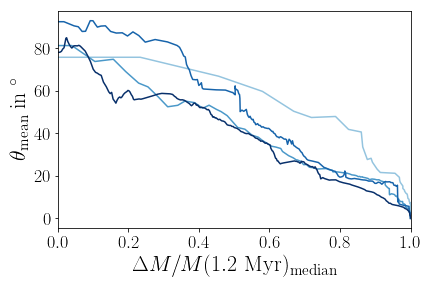}
\caption{Same as \Fig{t_thetamean_1p2M}, but for the mean of $\theta$ over median of relative mass accrual.}
\label{fig:dMmedian_thetamean_1p2M}
\end{figure}

\begin{figure}
\includegraphics[width=\linewidth]{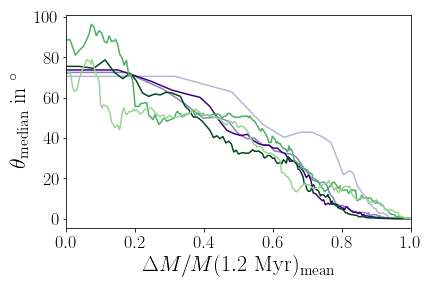}
\includegraphics[width=\linewidth]{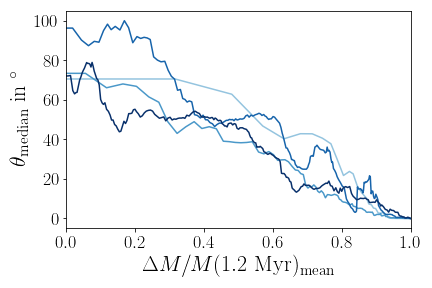}
\caption{Same as \Fig{t_thetamean_1p2M}, but for the median of $\theta$ over mean of relative mass accrual.}
\label{fig:dMmean_thetamedian_1p2M}
\end{figure}

\begin{figure}
\includegraphics[width=\linewidth]{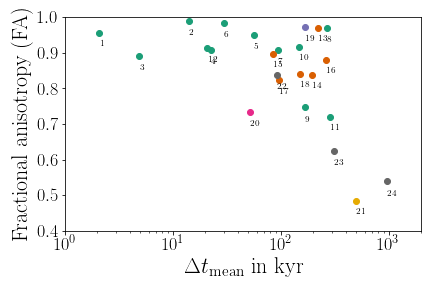}
\includegraphics[width=\linewidth]{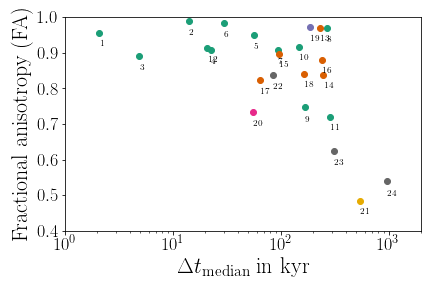}
\caption{Fractional anisotropy of multiple systems shown in \Fig{dt_FA}, but plotted over the mean (upper panel) and median (lower panel) of the difference in formation time of the members in the systems.
}\label{fig:dtmean_FA}

\end{figure}

\begin{figure}
\includegraphics[width=\linewidth]{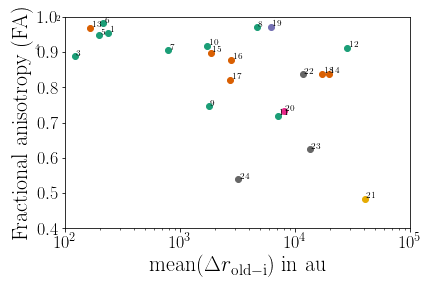}
\includegraphics[width=\linewidth]{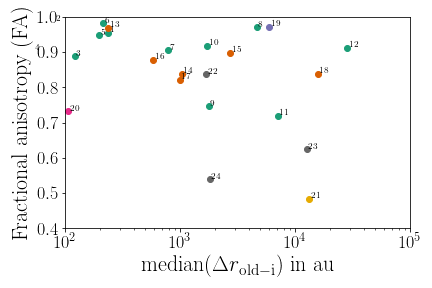}
\caption{Fractional anisotropy of multiple systems shown in \Fig{dt_FA}, but plotted over the mean (upper panel) and median (lower panel) of the distance to the oldest component.
}\label{fig:drmean_FA}

\end{figure}

\begin{figure}
\includegraphics[width=\linewidth]{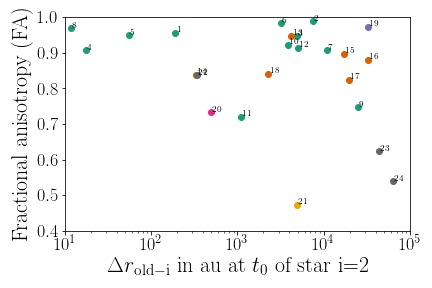}
\includegraphics[width=\linewidth]{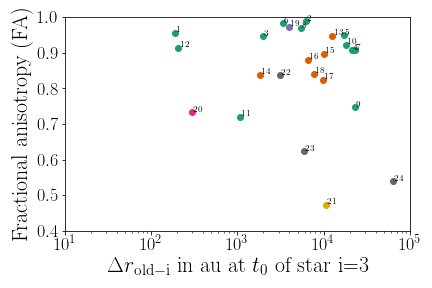}
\includegraphics[width=\linewidth]{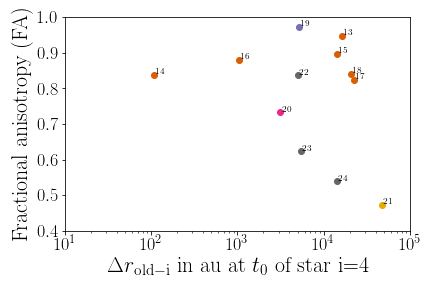}
\caption{Fractional anisotropy (FA) measured at $t=1.91$ Myr for the 24 systems shown in \Fig{dt_FA}, but plotted over distance between the oldest member ($1$) and member $i$ at the formation time of member $i$ from $i=2$ (second oldest star in the system; top panel) to $i=4$ (4th oldest star in the system; bottom panel). 
}\label{fig:dr_FA_old-1to3}

\end{figure}

\begin{figure}
\includegraphics[width=\linewidth]{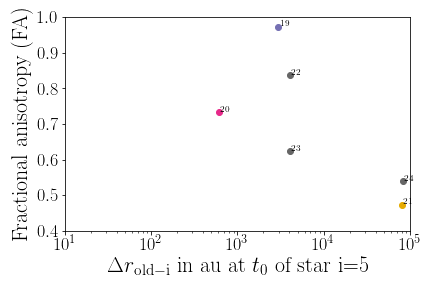}
\includegraphics[width=\linewidth]{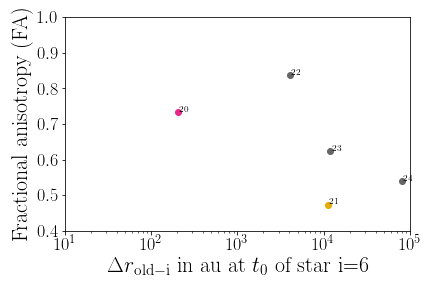}
\includegraphics[width=\linewidth]{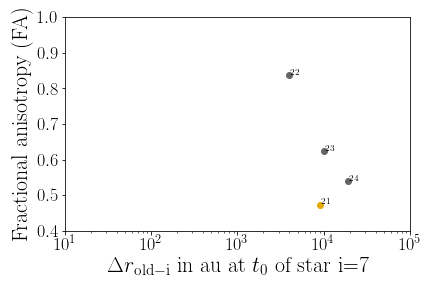}
\caption{Same as \Fig{dr_FA_old-1to3}, but for 5th to 7th member.
}\label{fig:dr_FA_old-4to6}

\end{figure}

\begin{figure}
\includegraphics[width=\linewidth]{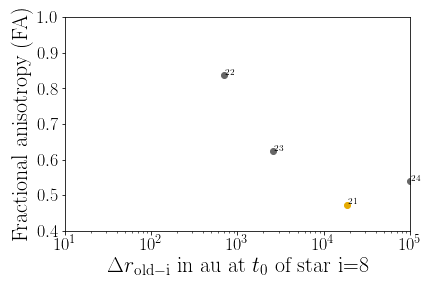}
\includegraphics[width=\linewidth]{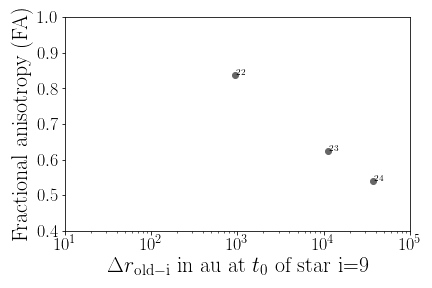}
\includegraphics[width=\linewidth]{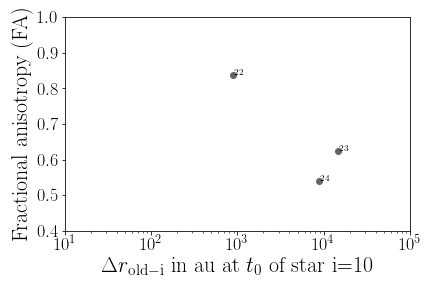}
\caption{Same as \Fig{dr_FA_old-1to3}, but for 8th to 10th member.
}\label{fig:dr_FA_old-7to9}

\end{figure}

\begin{figure}
\includegraphics[width=\linewidth]{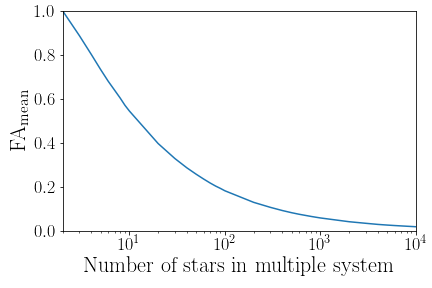}
\caption{Expectation value of fractional anisotropy FA for a random distribution of $n$ vectors. 
The expectation value is computed over 10000 iterations per system with $n$ vectors.
}\label{fig:n_FA_mean}

\end{figure}

Complementary to the plots show in the main part of the paper,
we also show the median of angle $\theta$ over time in \Fig{t_thetamedian_1p2M}, 
the median of angle $\theta$ over the median of relative mass accrual in \Fig{dMmedian_thetamedian_1p2M},
the mean of angle $\theta$ over the median of relative mass accrual in \Fig{dMmedian_thetamean_1p2M},
as well as the median of angle $\theta$ over the mean of relative mass accrual in \Fig{dMmean_thetamedian_1p2M}.
Moreover, we show fractional anisotropy over the mean and median of the difference in formation time between the system members in \Fig{dtmean_FA}, as well as FA over the mean and median of the difference in relative distance in \Fig{drmean_FA}. 
\Figure{dr_FA_old-1to3} to \Fig{dr_FA_old-7to9} show FA over the relative distance $\Delta r_{\rm old-i}$ with respect to the individual members. 
Finally, \Fig{n_FA_mean} shows the mean value of FA for multiple systems of n members based on a random distribution of 10000 iterations.  

\end{appendix}
\end{document}